\documentclass{article}

    \PassOptionsToPackage{numbers, compress}{natbib}

\usepackage[preprint]{neurips_2024}

\usepackage{amsmath,amsfonts,amssymb, amsthm}
\usepackage{array}
\usepackage{subfig}
\usepackage{caption}
\usepackage{colortbl}
\usepackage{textcomp}
\usepackage{stfloats}
\usepackage{url}
\usepackage{verbatim}
\usepackage{graphicx}
\usepackage{wrapfig}
\usepackage{bm}
\usepackage{enumitem}
\usepackage{multirow}
\usepackage{booktabs}       
\usepackage{flushend}
\usepackage[ruled,vlined]{algorithm2e}
\usepackage{bbding}
\usepackage{tikz,xcolor}
\usepackage[colorlinks,linkcolor=blue]{hyperref}
\usepackage{tcolorbox}

\newtheorem{proposition}{Proposition}

\newtheorem{lemma}{Lemma}
\newtheorem{theorem}{Theorem}

\newcommand{\ie}{\textit{i.e.}}
\newcommand{\eg}{\textit{e.g.}}

\newcommand{\improv}{\scalebox{1.2}{$\blacktriangle$\%} }
\newcommand{\gc}[1]{\textcolor{gray}{#1}}

\SetKwComment{tcp}{$\triangleright$ }{}

\title{
    Are LLM-based Recommenders Already the Best? Simple Scaled Cross-entropy Unleashes the Potential of Traditional Sequential Recommenders
}

\newcommand{\equalcontrib}{\textsuperscript{*}}

\author{%
    Cong Xu\equalcontrib \\
    East China Normal University \\
    Shanghai, China \\
    \texttt{congxueric@gmail.com} \\
    \And
    Zhangchi Zhu\equalcontrib \\
    East China Normal University \\
    Shanghai, China \\
    \texttt{zczhu@stu.ecnu.edu.cn} \\
    \And
    Mo Yu \\
    Pattern Recognition Center \\ 
    WeChat \\
    \texttt{moyumyu@global.tencent.com} \\
    \And
    Jun Wang \\
    East China Normal University \\
    Shanghai, China \\
    \texttt{wongjun@gmail.com} \\
    \And
    Jianyong Wang \\
    Tsinghua University \\
    Beijing, China \\
    \texttt{jianyong@tsinghua.edu.cn} \\
    \And
    Wei Zhang \\
    East China Normal University \\
    Shanghai, China \\
    \texttt{zhangwei.thu2011@gmail.com} \\
}

\everymath{\displaystyle} 
\begin{document}

\maketitle

\begin{abstract}

Large language models (LLMs) have been garnering increasing attention in the recommendation community.
Some studies have observed that LLMs, when fine-tuned by the cross-entropy (CE) loss with a full softmax,
could achieve `state-of-the-art' performance in sequential recommendation.
However, most of the baselines used for comparison are trained using a pointwise/pairwise loss function.
This inconsistent experimental setting leads to the underestimation of traditional methods 
and further fosters over-confidence in the ranking capability of LLMs.

In this study, we provide theoretical justification for the superiority of the cross-entropy loss 
by demonstrating its two desirable properties: tightness and coverage.
Furthermore, 
this study sheds light on additional novel insights:
1) Taking into account only the recommendation performance, 
CE is not yet optimal as it is not a quite tight bound in terms of some ranking metrics.
2) In scenarios that full softmax cannot be performed, 
an effective alternative is to scale up the sampled normalizing term.
These findings then help unleash the potential of traditional recommendation models, allowing them to surpass LLM-based counterparts.
Given the substantial computational burden,
existing LLM-based methods are not as effective as claimed for sequential recommendation.
We hope that these theoretical understandings in conjunction with the empirical results
will facilitate an objective evaluation of LLM-based recommendation in the future.
Our code is available at \url{https://github.com/MTandHJ/CE-SCE-LLMRec}.

\end{abstract}

\renewcommand{\thefootnote}{\fnsymbol{footnote}}
\footnotetext[1]{Equal contribution.}
\renewcommand{\thefootnote}{\arabic{footnote}}

\section{Introduction}

With the rapid growth of the Internet, 
the amount of information being generated every moment 
is far beyond human discernment.
Recommender systems are thus developed to help humans quickly and accurately ascertain the items of interest,
and have played increasingly important roles in diverse applications,
including e-commerce~\cite{zhou2018deep}, online news~\cite{gong2022positive}, and education~\cite{WuWZ24}.
Due to inherent differences in data types and recommendation goals, 
different tasks are typically handled using various techniques and models.
For example, graph neural networks \cite{GNN:GCN:Kipf:2016} have become predominant in collaborative filtering \cite{CF:LightGCN:He:2020,CF:UltraGCN:Mao:2021},
while Transformer \cite{Transformer:Attention:Vaswani:2017} has gained increasing popularity in sequential recommendation~\cite{Seq:SASRec:Kang:2018,Seq:Bert4Rec:Sun:2019}.

Recently, the prosperity of Large Language Models (LLMs) \cite{LLM:T5:Raffel:2020,LLM:GPT:Openai:2023,LLM:LLama:Touvron:2023,LLM:LLaMA2:Touvron:2023} 
suggests a promising direction towards universal recommenders~\cite{LLM4Rec:P5:Geng:2022,LLM4Rec:POD:Li2023}.
Equipped with carefully designed prompts, 
they show great potential in explainable and cross-domain recommendations \cite{LLM4Rec:CRs:Feng:2023,LLM4Rec:Chat-Rec:Gao:2023}.
Nevertheless, 
there remain non-negligible gaps \cite{LLM4Rec:Prefrence:Kang:2023,LLM4Rec:TallRec:Bao:2023} between LLMs and traditional methods
unless domain-specific knowledge is injected.
After fine-tuning~\cite{LLM4Rec:E4SRec:Li:2023,LLM4Rec:LLaMARec:Yue2023},
some researches have observed `compelling' results 
and hastily affirmed LLM-based recommenders' ranking capability.
Considering the substantial computational burden that comes with LLMs, 
we contend that such claims must be carefully scrutinized before being accepted and applied, 
to prevent a massive waste of resources.
Recall that the next-token prediction objective used for LLM pre-training (and fine-tuning)
inherently relies on a cross-entropy (CE) loss that necessitates a full softmax over the entire corpus. 
In contrast, most traditional recommenders are trained using pointwise/pairwise losses;
for example, either BCE adopted in SASRec~\cite{Seq:SASRec:Kang:2018} or BPR in FMLP-Rec~\cite{Seq:FMLPRec:Zhou:2022} dynamically samples one negative item for each positive item.
Some studies \cite{loss:NEG4Rec:Chen:2023,loss:NEG4Rec:Klenitskiy:2023} have pointed out that CE typically outperforms BCE or BPR in sequential recommendation, 
which raises our concerns about the fairness of previous comparisons.
In this paper, we are to clarify these concerns holistically.

\textbf{Q1. Do LLM-based recommenders outperform traditional models when the cross-entropy loss is applied to both?}
Since non-ID models tend to have lower performance compared to ID-integrated methods,
we conducted a comparison between four state-of-the-art LLM-based recommenders with an independent ID embedding table
and four traditional models upon MLP, CNN, RNN, or Transformer.
Interestingly,
although traditional methods do fall far behind when using BCE or BPR,
some of them outperform LLMs by a large margin once the cross-entropy loss is applied.

\textbf{Q2. How to understand the superiority of cross-entropy loss in improving the ranking capability of recommenders?}
On the one hand,
Bruch et al.~\cite{loss:Bound:Bruch:2019} has theoretically established a connection between the cross-entropy loss and some ranking metrics.
On the other hand, Wu et al.~\cite{loss:SSM:Wu:2023} further proved its beneficial property in alleviating popularity bias.
Here, we follow up on \cite{loss:Bound:Bruch:2019} to reveal further novel insights: 
an ideal recommendation loss should emphasize both \textit{tightness} and \textit{coverage}.
The former necessitates that the recommendation loss should be designed to be a good soft proxy of the ranking metrics
for consistent improvements,
while the latter underscores the exploration of an adequate number of negative samples during the training process.
Clearly, CE is optimal for coverage as it incorporates all items during each mini-batch iteration. 
However, we theoretically demonstrate that it remains suboptimal in tightness.
Superior recommendation performance can be obtained through a dynamic truncation.

\textbf{Q3. Can traditional models still outperform LLM-based recommenders when full softmax cannot be performed?}
It is important to note that in practical scenarios, the number of items can reach hundreds of millions, 
making it infeasible to perform full softmax accurately.
Thanks to the advent of subword segmentation algorithms \cite{Subword:Unigram:Kudo:2018},
the corpus size of a language model is constrained, 
thereby providing a feasible solution to the aforementioned problem 
as long as the separate IDs can be represented in the same manner \cite{LLM4Rec:CIDIID:Hua2023}.
Consequently, 
if traditional models cannot fully unleash their potential in this practical situation,
current LLM-based recommenders' performance remains highly promising.
Inspired by the theoretical findings presented in Q2, 
we propose a very simple alternative named Scaled Cross-Entropy (SCE) to address the concerns in reality. 
SCE uniformly samples items to estimate the normalization term of cross-entropy.
In contrast to the normal sampled softmax loss that exhibits poor `tightness' (as formally justified in Section~\ref{section-sce}),
SCE addresses this issue by scaling up the sampled term using an additional weight.
As a result, sampling a very small number of negative samples per iteration is sufficient to achieve comparable results with using cross-entropy by a full softmax.

The remainder of this paper is organized as follows. 
A preliminary investigation concerning their performance will be conducted in Section~\ref{section-observation}.
Then Section~\ref{section-tightness-coverage} will discuss Q2 in theory, while a practical comparison for Q3 will be found in Section~\ref{section-sce}.
We finally review related work on sequential recommendation and LLM-based recommendation in Section~\ref{section-related-work}.

\section{Potential of Traditional Recommenders over LLM-based Recommenders}
\label{section-observation}

In this section, we showcase how the performance of the traditional sequential recommendation models changes when switching from BCE/BPR to CE.
The experimental setup will be introduced first.

\begin{table*}[]
    \centering
    \caption{
    Loss functions discussed in this paper and their bounding probabilities in the case of uniform sampling.
    $\sigma(\cdot)$ stands for the sigmoid function.
    }
    \label{table-summary-loss}
\scalebox{0.82}{
\begin{tabular}{l||c|c|c}
    \toprule
    Loss  
    & Formulation 
    & `Normalizing' term $Z$ 
    & Bounding probability $\ge$
    \\
    \midrule
    $\ell_{\text{CE}}$
    & $ -\log \frac{\exp(s_{v_+})}{ \sum_{v \in \mathcal{I}} \exp(s_v)}$ 
    & $ \sum_{v \in \mathcal{I}} \exp(s_v)$
    & $1$
    \\
    \midrule
    $\ell_{\text{BCE}}$
    & $ -\log \sigma(s_{v_+}) - \log (1 - \sigma(s_{v_-}))$ 
    & $ (1 + \exp(s_{v_+})) (1 + \exp(s_{v_-}))$
    & 0
    \\
    $\ell_{\text{BPR}}$
    & $ -\log \sigma(s_{v_+} - s_{v_-})$ 
    & $ \exp(s_{v_+}) + \exp(s_{v_-})$
    & 0
    \\
    \midrule
    $\ell_{\text{SSM}}$
    & $ -\log \frac{\exp(s_{v_+})}{\exp(s_{v_+}) + 1 \cdot \sum_{i=1}^K \exp(s_{v_i})} $ 
    & $ \exp(s_{v_+}) + \sum_{i=1}^K \exp(s_{v_i}) $
    & $1 - m \big(1 - r_+ / |\mathcal{I}| \big)^{\lfloor K / m \rfloor}$
    \\
    $\ell_{\text{SCE}}$
    & $ -\log \frac{\exp(s_{v_+})}{\exp(s_{v_+}) + \alpha \sum_{i=1}^K \exp(s_{v_i})} $ 
    & $ \exp(s_{v_+}) + \alpha \sum_{i=1}^K \exp(s_{v_i}) $
    & $1 -\lceil\frac{m}{\alpha} \rceil \big(1 - r_+ / |\mathcal{I}| \big)^{\lfloor K / \lceil \frac{m}{\alpha} \rceil \rfloor}$
    \\
    \bottomrule
\end{tabular}
}
\end{table*}

\textbf{Problem definition.} Given a query $q$ that encompasses some user information,
a recommender system aims to retrieve some items $v \in \mathcal{I}$ that would be of interest to the user.
In sequential recommendation, 
it is to predict the next item $v_{t+1}$ based on historical interactions $q = [v_1, v_2, \ldots, v_t]$.
The crucial component is to develop a scoring function $s_{qv} := s_{\theta}(q, v)$ to accurately model the relevance of a query $q$ to a candidate item $v$.
A common paradigm is to map them into the same latent space through some models parameterized via $\theta$, 
followed by an inner product operation for similarity calculation.
Then, top-ranked items based on these scores will be prioritized for recommendation.
Typically the desired recommender is trained to minimize an optimation objective over all observed interactions $\mathcal{D}$: 
\begin{equation*}
    \min_{\theta} \quad \mathbb{E}_{(q, v_+) \sim \mathcal{D}} [\ell(q, v_+; \theta)],
\end{equation*}
where $v_+$ indicates the target item for the query $q$, 
and the loss function $\ell$ considered in this paper (see Table \ref{table-summary-loss}) can be reformulated as
\begin{equation}
    \label{eq-loss-unified}
    \ell(q, v_+; \theta) := -\log \frac{
        \exp(s_{\theta}(q, v_+))
    }{
        Z_{\theta}(q)
    }
    = -s_{\theta}(q, v_+) + \log Z_{\theta}(q).
\end{equation}
Here, we specifically separate the `normalizing' term $Z_{\theta}(q)$, 
as it plays an important role in determining the connection to ranking metrics.
We will omit $q$ and $\theta$ hereafter if no ambiguity is raised.

\textbf{Datasets.}
We select two public datasets, \ie,  Beauty and Yelp,
which have been widely used in previous studies~\cite{Seq:SASRec:Kang:2018,LLM4Rec:P5:Geng:2022,LLM4Rec:POD:Li2023}.
Following the studies~\cite{Seq:FMLPRec:Zhou:2022,LLM4Rec:P5:Geng:2022},
we filter out users and items with less than 5 interactions,
and the validation set and test set are split in a \textit{leave-one-out} fashion, 
namely the last interaction for testing and the penultimate one for validation.
The dataset statistics after pre-processing are presented in Appendix~\ref{section-dataset}.

\textbf{Evaluation metrics.}
For each user, the scores returned by the recommender will be sorted in descending order to generate candidate lists. 
Denoted by $r_+ := r(v_+) = |\{v \in \mathcal{I}: s_v \ge s_{v_+}\}|$ the predicted rank of the target item $v_+$,
Normalized Discounted Cumulative Gain (NDCG) and Mean Reciprocal Rank (MRR)
\footnote{
In practice, it is deemed meaningless when $r_+$ exceeds a pre-specified threshold $k$ (\eg, $k=1,5,10$).
Hence, the widely adopted NDCG@$k$ and MRR@$k$ metrics are modified to assign zero rewards to these poor ranking results.
}
are often employed to assess the sorting quality.
For next-item recommendation considered in this paper, NDCG and MRR (for a single query) can be simplified to
\begin{equation}
    \text{NDCG}(r_+) = \frac{1}{\log_2 (1 + r_+)}, 
    \quad \text{MRR}(r_+) = \frac{1}{r_+}.
\end{equation}
Both of them increase as the target item $v_+$ is ranked higher and reaches the maximum when $v_+$ is ranked first (\ie, $r_+ = 1$).
Consequently, 
the average quality computed over the entire test set serves as an indicator of the ranking capability.

\textbf{Baselines.}
Although LLM itself has surprising zero-shot recommendation ability,
there still exist non-negligible gaps \cite{LLM4Rec:Prefrence:Kang:2023,LLM4Rec:TallRec:Bao:2023} unless domain-specific knowledge is injected.
Hence, only LLM-based recommenders enhanced by an independent ID embedding table will be compared in this paper:
P5 (CID+IID)~\cite{LLM4Rec:CIDIID:Hua2023}, POD~\cite{LLM4Rec:POD:Li2023},
LlamaRec~\cite{LLM4Rec:LLaMARec:Yue2023}, and E4SRec~\cite{LLM4Rec:E4SRec:Li:2023}.
Specifically, the first two methods take T5 as the foundation model,
while the last two methods fine-tune Llama2 for efficient sequential recommendation.
Because the data preprocessing scripts provided with POD may lead to information leakage \cite{LLM4Rec:Tiger:Rajput:2023},
we assign random integer IDs to items rather than sequentially incrementing integer IDs.
Additionally, 
four sequence models including 
Caser \cite{Seq:Caser:Tang:2018}, 
GRU4Rec \cite{Seq:GRU4Rec:Hidasi:2015}, 
SASRec~\cite{Seq:SASRec:Kang:2018}, 
and FMLP-Rec \cite{Seq:FMLPRec:Zhou:2022}
are considered here to unveil the true capability of traditional methods.
They cover various architectures so as to comprehensively validate the effectiveness of the proposed approximation methods.
Table~\ref{table-model-statistics} presents an overview of the model statistics.
Other implementation details can be found in Appendix~\ref{section-settings}.

\begin{table}[!t]
    \centering
    \caption{
        Model statistics.
        'Max Seq. Len.' denotes the maximum input sequence length while
        `Max ID Seq. Len.' indicates the maximum number of historical user interactions used for prediction.
    }
    \label{table-model-statistics}
    \scalebox{0.75}{
\begin{tabular}{lcccccc}
    \toprule
\multicolumn{1}{c}{Model} & Foundation Model & Architecture & Emb. Size & Max Seq. Len. & Max ID Seq. Len. & \#Params \\
    \midrule
P5 (CID+IID) \cite{LLM4Rec:CIDIID:Hua2023}                               & T5                                & Transformer                   & 512                        & 512                            & 50                                & 60M                     \\
POD \cite{LLM4Rec:POD:Li2023}                                        & T5                                & Transformer                   & 512                        & 512                            & 50                                & 60M                     \\
LlamaRec \cite{LLM4Rec:LLaMARec:Yue2023}                                   & Llama2                            & Transformer                   & 4096                       & 2048                           & 50                                & 7B                      \\
E4SRec \cite{LLM4Rec:E4SRec:Li:2023}                                     & Llama2                            & Transformer                   & 4096                       & 2048                           & 50                                & 7B                      \\
    \midrule
GRU4Rec \cite{Seq:GRU4Rec:Hidasi:2015}                                    &                                   & RNN                           & 64                         & 50                             & 50                                & 0.80M                   \\
Caser \cite{Seq:Caser:Tang:2018}                                      &                                   & CNN                           & 64                         & 50                             & 50                                & 3.80M                   \\
SASRec \cite{Seq:SASRec:Kang:2018}                                     &                                   & Transformer                   & 64                         & 50                             & 50                                & 0.83M                   \\
FMLP-Rec \cite{Seq:FMLPRec:Zhou:2022}                                  &                                   & MLP                           & 64                         & 50                             & 50                                & 0.92M                   \\
    \bottomrule
\end{tabular}
    }
\end{table}
\begin{table}[!t]
    \centering
    \caption{
        Overall performance comparison on the Beauty and Yelp datasets.
        The best results of each block are marked in \textbf{bold}.
        \gc{Gray} results indicate less than that of best LLM-based recommenders.
        `\improv over LLM' represents the relative gap between respective best results.
    }
    \label{table-overall-bpr-bce-ce}
    \scalebox{0.72}{
\begin{tabular}{c|l|cccc|cccc}
    \toprule
\multicolumn{2}{c|}{\multirow{2}{*}{Method}}               & \multicolumn{4}{c}{Beauty}                                                                                                                                        & \multicolumn{4}{c}{Yelp}                                                                                                      \\

\multicolumn{1}{c}{}  &          & NDCG@5                                 & NDCG@10                                & MRR@5                                  & MRR@10                                 & NDCG@5                        & NDCG@10                       & MRR@5                         & MRR@10                        \\
    \midrule
    \midrule
     \multirow{4}{*}{LLM}                  & POD         & 0.0125                                 & 0.0146                                 & 0.0107                                 & 0.0116                                 & \textbf{0.0330}                        & 0.0358                        & 0.0267                        & 0.0292                        \\
                       & P5(CID+IID) & 0.0403                                 & 0.0474                                 & \textbf{0.0345}                        & 0.0376                                 & 0.0200                        & 0.0252                        & 0.0164                        & 0.0186                        \\
                       & LlamaRec    & \textbf{0.0405}                        & \textbf{0.0492}                        & 0.0344                                 & \textbf{0.0380}                        & 0.0306               & \textbf{0.0367}               & \textbf{0.0270}               & \textbf{0.0295}               \\
  & E4SRec      & 0.0376                                 & 0.0448                                 & 0.0326                                 & 0.0356                                 & 0.0207                        & 0.0260                        & 0.0174                        & 0.0196                        \\

    \midrule
    \midrule
 \multirow{4}{*}{BPR}                      & GRU4Rec     & \gc{0.0155}          & \gc{0.0208}          & \gc{0.0125}          & \gc{0.0147}          & \gc{0.0106} & \gc{0.0147} & \gc{0.0086} & \gc{0.0102} \\
                       & Caser       & \gc{0.0173}          & \gc{0.0221}          & \gc{0.0143}          & \gc{0.0163}          & \gc{0.0211} & \gc{0.0250} & \gc{0.0185} & \gc{0.0201} \\
                       & SASRec      & \gc{0.0298}          & \gc{0.0374}          & \gc{\textbf{0.0248}} & \gc{0.0279}          & \gc{0.0167} & \gc{0.0218} & \gc{0.0137} & \gc{0.0158} \\
  & FMLP-Rec    & \gc{\textbf{0.0321}} & \gc{\textbf{0.0393}} & \gc{0.0269}          & \gc{\textbf{0.0298}} & \textbf{0.0334}               & \textbf{0.0405}               & \textbf{0.0286}               & \textbf{0.0314}               \\
\multicolumn{2}{c|}{\improv over LLM} & -20.8\%                                & -20.3\%                                & -22.1\%                                & -21.5\%                                & +1.3\%                         & +10.3\%                        & +5.7\%                         & +6.5\%                         \\
    \midrule
 \multirow{4}{*}{BCE}                      & GRU4Rec     & \gc{0.0134}          & \gc{0.0186}          & \gc{0.0108}          & \gc{0.0129}          & \gc{0.0098} & \gc{0.0135} & \gc{0.0079} & \gc{0.0094} \\
                       & Caser       & \gc{0.0185}          & \gc{0.0234}          & \gc{0.0154}          & \gc{0.0174}          & \gc{0.0224} & \gc{0.0264} & \gc{0.0198} & \gc{0.0214} \\
                       & SASRec      & \gc{0.0275}          & \gc{0.0353}          & \gc{0.0225}          & \gc{0.0257}          & \gc{0.0225} & \gc{0.0281} & \gc{0.0192} & \gc{0.0215} \\
  & FMLP-Rec    & \gc{\textbf{0.0301}} & \gc{\textbf{0.0381}} & \gc{\textbf{0.0248}} & \gc{\textbf{0.0281}} & \gc{\textbf{0.0330}}               & \textbf{0.0391}               & \textbf{0.0287}               & \textbf{0.0312}               \\
\multicolumn{2}{c|}{\improv over LLM} & -25.8\%                                & -22.6\%                                & -28.1\%                                & -26.0\%                                & -0.1\%                        & +6.5\%                         & +6.1\%                         & +5.7\%                         \\
    \midrule
    \midrule
 \multirow{4}{*}{CE}                      & GRU4Rec     & \gc{0.0329}          & \gc{0.0398}          & \gc{0.0281}          & \gc{0.0310}          & \gc{0.0171} & \gc{0.0231} & \gc{0.0137} & \gc{0.0161} \\
                       & Caser       & \gc{0.0303}          & \gc{0.0361}          & \gc{0.0260}          & \gc{0.0284}          & \gc{0.0211} & \gc{0.0243} & \gc{0.0187} & \gc{0.0200} \\
                       & SASRec      & \textbf{0.0510}                        & \textbf{0.0597}                        & \textbf{0.0443}                        & \textbf{0.0479}                        & 0.0345                        & 0.0415                        & 0.0302                        & 0.0331                        \\
   & FMLP-Rec    & 0.0507                                 & 0.0594                                 & 0.0438                                 & 0.0473                                 & \textbf{0.0364}               & \textbf{0.0444}               & \textbf{0.0316}               & \textbf{0.0348}               \\
\multicolumn{2}{c|}{\improv over LLM} & +25.7\%                                 & +21.3\%                                 & +28.1\%                                 & +26.0\%                                 & +10.3\%                        & +21.0\%                        & +16.9\%                        & +18.1\%                       \\
    \bottomrule
\end{tabular}
    }
\end{table}

\textbf{Observations.}
Let us focus on the first three blocks in Table \ref{table-overall-bpr-bce-ce},
where the majority of traditional models notably fall behind LLM-based recommenders (\eg, LlamaRec).
Although the current state-of-the-art model, FMLP-Rec, achieves fairly good results on Yelp, 
it still does not perform as well on Beauty compared to P5, LlamaRec, and E4SRec, 
with a performance gap of at least 20\%.
In particular, 
SASRec being a Transformer-based model similar to LLM-based recommenders,
unfortunately performs poorly in comparison.
It is tempting to attribute this result to the superior model expressive power and world knowledge of LLMs. 
However, this does not seem to be the case based on the results obtained using CE:
SASRec and FMLP-Rec demonstrate significant performance improvements as a result of this subtle change, 
ultimately surpassing all LLM-based recommenders on both the Beauty and Yelp datasets.
GRU4Rec and Caser also benefit from using CE, although they may not be able to outperform LLM-based recommenders due to their limited expressive power.

Hence, we argue that previous affirmative assertions about the LLMs' recommendation performance 
are rooted in biased comparisons, wherein BCE or BPR are commonly used for training traditional models.
Moreover, the inflated model size (from 60M of P5 to 7B of LlamaRec) only yields marginal improvements in some of the metrics.
The rich world knowledge and powerful reasoning ability seem to be of limited use here.
In conclusion, even after fine-tuning, 
LLM-based recommenders fail to outperform state-of-the-art traditional sequence models when full softmax can be performed.

\section{Tightness and Coverage for Normalizing Term}
\label{section-tightness-coverage}

In this section, we are to clarify the superiority of the cross-entropy loss by examining the desirable properties of its normalizing term.
This analysis provides some valuable insights into the design of a recommendation loss that is able to boost the ranking capability.
SASRec \cite{Seq:SASRec:Kang:2018} as the most prominent sequence models
will serve as the baseline to empirically elucidate the conclusions in this part.
All results are reported over 5 independent runs.

\subsection{Tightness}
To accurately predict the next item a user is likely to purchase, 
the model must have superior ranking capability,
which is quantitatively assessed by the aforementioned metrics NDCG and MRR.
Ideally, the recommender should be optimized directly towards these metrics;
however, this cannot be achieved exactly due to their discrete nature.
If a recommendation loss can be regarded as a soft proxy for NDCG or MRR, 
then minimizing this loss is expected to consistently improve ranking metrics as well.
Bruch et al. \cite{loss:Bound:Bruch:2019} has theoretically connected the cross-entropy loss to these ranking metrics, 
as stated below:

\begin{lemma}[\cite{loss:Bound:Bruch:2019}]
    \label{lemma-ce-ndcg-mrr}
    Minimizing the cross-entropy loss $\ell_{\mathrm{CE}}$ is equivalent to maximizing a lower bound of NDCG and MRR.
\end{lemma}
Indeed, Lemma~\ref{lemma-ce-ndcg-mrr} can be extended to a more generalized version, 
which can provide us with a technical foundation for further investigations:
\begin{proposition}
    \label{proposition-bounding}
    For a target item $v_+$ which is ranked as $r_+$, the following inequality holds true for any $n \ge r_+$
    \begin{equation}
        -\log \mathrm{NDCG}(r_+) \le -\log \mathrm{MRR}(r_+) \le \ell_{\mathrm{CE}\text{-}n}.
    \end{equation}
    where
    \begin{equation}
        \ell_{\mathrm{CE}\text{-}n} := -s_{v_+}  + \log \sum_{r(v) \le n} \exp(s_{v}).
    \end{equation}
\end{proposition}
\begin{proof}
    Proof of Proposition~\ref{proposition-bounding} is in Appendix~\ref{section-proof-prop1}.
\end{proof}

Proposition \ref{proposition-bounding} implies that all CE-like losses $\ell_{\mathrm{CE}\text{-}n}$ 
possess the potential to act as a soft proxy for NDCG and MRR, 
provided that all items ranked before $v_+$ are retained in the normalizing term.
Moreover, the connection becomes tighter as $n$ becomes smaller.
From this point of view, the original cross-entropy loss appears to be the `worst' since it is a special case that includes all items (\ie, $n=|\mathcal{I}|$).
We further validate that tightness is not the sole principle of worth, and it comes at the expense of certain coverage.

\subsection{Coverage}
\begin{figure}
	\centering
    \subfloat[$\eta$-ablation]{
        \label{fig-tighter-bounds}
        \includegraphics[width=0.6\textwidth]{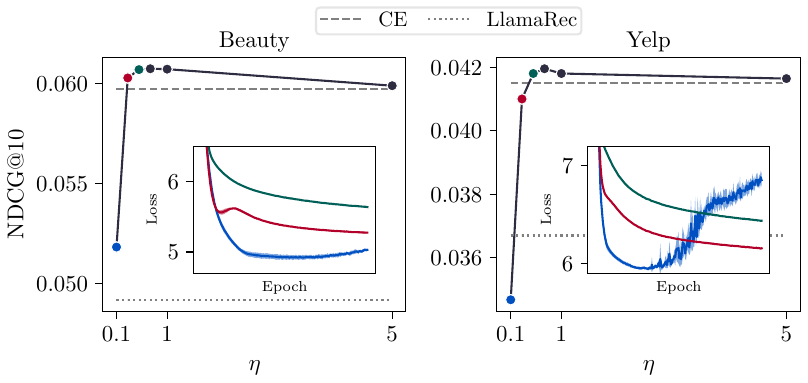}
    }
    \subfloat[Tightness versus Coverage]{
        \label{fig-tightness-coverage}
        \includegraphics[width=0.3\textwidth]{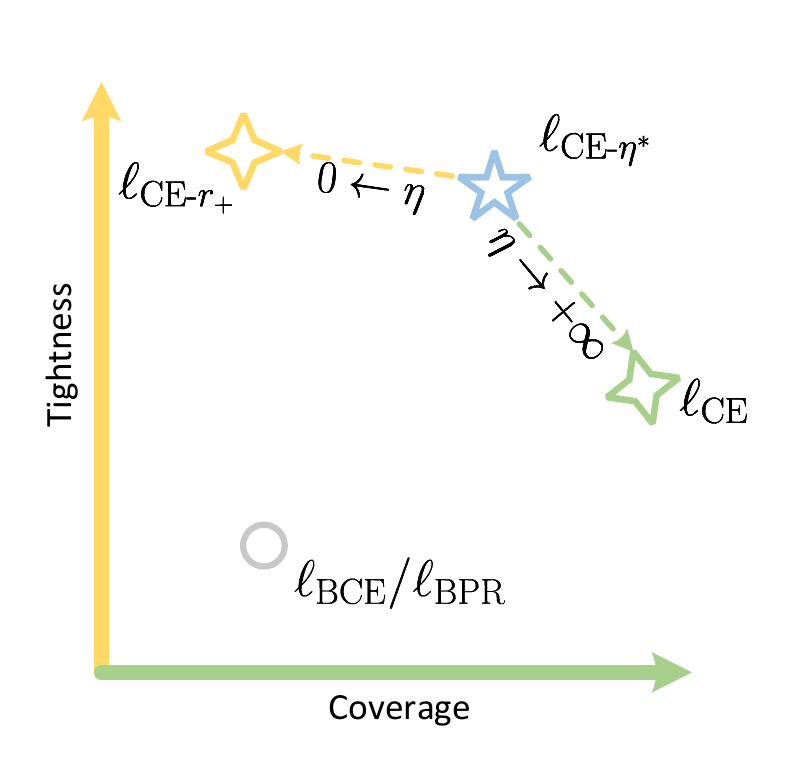}
    }
	
	\caption{
        (a) Performance comparison based on tighter bounds for NDCG.
        The dashed line represents the results trained by CE (namely the case of $\eta \rightarrow +\infty$).
        (b) Tightness and Coverage illustration of different loss functions.
        $\eta^*$ is a task-specific value derived from (a).
	}
\end{figure}

Since the superiority of CE possibly stems from its connection to these ranking metrics, 
one might hypothesize that optimizing a tighter bound with a smaller value of $n \ll |\mathcal{I}|$ allows greater performance gains.
However, the condition $n \ge r_+$ in Proposition~\ref{proposition-bounding} cannot be consistently satisfied via a constant value of $n$ 
since $r_+$ is dynamically changing during training.
Alternatively, an adaptive truncation can be employed for this purpose:

\begin{equation}
    \label{eq-truncation}
    \ell_{\text{CE-}\eta} := -s_{v_+} + \log \sum_{s_v - s_+ \ge -\eta |s_+|} \exp(s_{v}), \quad \eta \ge 0.
\end{equation}
Note that this \textit{$\eta$-truncated} loss retains only items whose scores are not lower than $s_+ - \eta |s_+|$,
so a tighter bound will be obtained as $\eta$ drops to 0.
Specifically, this $\eta$-truncated loss becomes $\ell_{\text{CE-}r_+}$ (\ie, the tightest case) when $\eta = 0$, and approaches $\ell_{\text{CE}}$ when $\eta \rightarrow +\infty$.
Figure~\ref{fig-tighter-bounds} illustrates how NDCG@10 varies as $\eta$ gradually increases from 0.1 to 5.
There are two key observations:
\begin{itemize}[leftmargin=*]
    \item[1.]
    SASRec performs worst in the tightest case of $\eta \approx 0$.
    This can be attributed to the poor coverage of the $\eta$-truncated loss:
    due to the truncation operation, very rare negative items are encountered during training.
    Moreover, this situation is exacerbated for targets that are easy to recognize, as they get high scores only after a few iterations.
    Consequently over-fitting is more likely to occur, resulting in deteriorated performance. 
    Nevertheless, all CE-like losses demonstrate competitive performance compared to LLM-based recommenders, 
    further underscoring the importance of tightness for a recommendation loss.
    \item[2.]
    Once $\eta$ is large enough to exceed the necessary coverage threshold,
    SASRec begins to benefit from tightness and achieves its best performance around $\eta \approx 0.7$.
    Further increasing $\eta$ however leads to a similar effect as the original cross-entropy loss, thereby along with a slightly degenerate performance due to the suboptimal tightness.
\end{itemize}

\textbf{Conclusions.}
We summarize the characteristics of tightness and coverage for different losses in Figure \ref{fig-tightness-coverage}.
$\ell_{\text{CE-}r_+}$ and $\ell_{\text{CE}}$ represent two extremes in terms of tightness and coverage, respectively.
However, neither of them is optimal when taking into account only the ranking capability.
There actually exists a sweet spot $\ell_{\text{CE-}\eta^*}$ whose tightness and coverage are well balanced.
These recommendation loss design principles 
have been indirectly employed in previous studies \cite{loss:IS4Rec:Lian:2020,loss:IS4Rec:Chen:2022,loss:Sampling4Rec:Shi:2023}:
while they focus on exploiting hard samples (\textit{for tightness}),
these items are still drawn from the entire pool of items (\textit{for coverage}) rather than from a fixed subset.
From this point of view, BCE and BPR are far from mature.
On the one hand, BCE and BPR sample one negative item for each query, which greatly restricts the coverage.
In practice, 
even after training for thousands of epochs, 
there is no guarantee that a user will encounter all the negative items, 
especially when the number of items is large.
On the other hand, 
consuming more training iterations may not alleviate this problem
because of their poor tightness (unless given stronger assumptions, no meaningful bounds can be derived for BCE or BPR).
Due to the large item size,
the negative sample uniformly drawn from them is often too easy to update the model in a meaningful direction.

\section{Scaled Cross-Entropy for Practical Potential of Traditional Models}
\label{section-sce}

In practical scenarios, the number of items can reach hundreds of millions, 
possibly making a precise execution of full softmax infeasible. 
It is noteworthy that 
despite the formulation of $\ell_{\text{CE-}n}$ or $\ell_{\text{CE-}\eta}$ 
involving only a limited number of items, 
a full softmax operation is still required before truncation.
Therefore, it may not be advisable to perform the aforementioned losses directly in real-world situations.
Conversely, LLM-based recommenders may effectively address this problem 
based on the commonly used subword segmentation algorithms \cite{Subword:Unigram:Kudo:2018} that allow for a fixed corpus.
As long as no independent ID embedding table\footnote{
    While non-ID recommenders are appealing and hold significant potential, there remains a substantial performance disparity compared to ID-based methods.
    Hua et al. \cite{LLM4Rec:CIDIID:Hua2023} have comprehensively investigated various item indexing methods,
    demonstrating that P5 enhanced by CID (collaborative ID) and IID (independent ID) significantly outperforms other approaches.
    This is the reason why we solely consider P5 (CID+IID) in this paper.
}
is introduced,
the full softmax cost then can be controlled at a constant value regardless of the number of items.
Consequently, 
in scenarios where traditional models fail to fully unleash their potential,
current LLM-based recommenders' performance remains highly promising.
Recall that concurrent with the advent of subword segmentation algorithms,
some studies resorted to approximations as substitutes for CE.
For example, Mnih et al.~\cite{loss:NCE4LM:Mnih:2012} simplified noise contrastive estimation 
for bypassing an explicit normalizing over the entire vocabulary.
Similarly, we turn to the sampled softmax loss for efficiency.

\begin{figure}[t]
	\centering
	\includegraphics[width=0.95\textwidth]{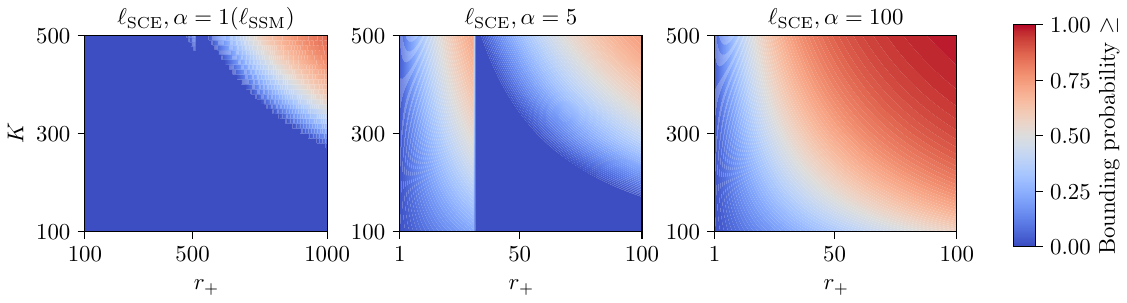}
	\caption{
        Bounding probabilities of SCE at different values of $\alpha = 1, 5, 100$ across varying ranks $r_+$ and number of negative items $K$. 
        The probabilities are calculated based on Eq.~\eqref{eq-ssm-bound-prob} and Eq.~\eqref{eq-sce-bounds}, with negative values being clipped to zero.
        For the Beauty dataset, $|\mathcal{I}| = 12101$.
	}
	\label{fig-bound-probs}
\end{figure}

Based on the findings in Section~\ref{section-tightness-coverage},
a desirable approximation to CE should exhibit not only tightness but also coverage.
Since the normalizing term of cross-entropy is intractable in reality,
a straightforward way is to approximate it by (uniformly) sampling part of items from $\mathcal{I}$:
\begin{equation}
    \ell_{\text{SSM}} := -s_{v_+} + \log \hat{Z}, \quad 
    \hat{Z} = \exp(s_{v_+}) + \sum_{i=1}^K \exp(s_{v_i}),
\end{equation}
which yields the well-known sampled softmax loss (SSM)~\cite{loss:SSM:Wu:2023,loss:NEG4Rec:Klenitskiy:2023}.
It is worth noting that SSM can have different variants depending on the scoring functions and sampling strategies employed.
Although cosine similarity and adaptive sampling probability~\cite{loss:IS4Rec:Lian:2020,loss:IS4Rec:Chen:2022} are expected to yield slightly better performance, 
the resulting additional memory and computational costs compromise efficiency.
Hence, the sampled softmax loss considered here is limited to the use of the inner product scoring function and uniform sampling
due to their widespread adoption and efficient implementation in recommender systems.
Despite the efficiency gained from uniform sampling,
there exists a very particular trade-off between tightness, coverage, and efficiency when it comes to this sampled loss. 
Firstly, coverage and efficiency are in conflict with each other.
Secondly, the discussion of tightness is not very rigorous for $\ell_{\text{SSM}}$ 
since it has a low probability of bounding NDCG in most cases:

\begin{theorem}
    \label{theorem-ssm-bounds}
    Let $v_+$ be a target item which is ranked as $r_+ \le 2^m - 1$ for some $m \in \mathbb{N}$,
    If we uniformly sample $K$ items for training, then with a probability of at least
    \begin{equation}
        \label{eq-ssm-bound-prob}
        1 - m\big(1 - r_+ / |\mathcal{I}| \big)^{\lfloor K / m \rfloor},
    \end{equation}
    we have
    \begin{equation}
        \label{ieq-ndcg-mrr}
        -\log \mathrm{NDCG}(r_+) \le  \ell_{\mathrm{SSM}}.
    \end{equation}
\end{theorem}
\begin{proof}
    Proof of Theorem~\ref{theorem-ssm-bounds} and corresponding results of MRR are in Appendix~\ref{section-proof-theorem12}.
\end{proof}

According to the estimates of the bounding probability in Eq.~\eqref{eq-ssm-bound-prob},
we illustrate the distribution of probability values for varying $r_+$ and $K$ in the left panel of Figure~\ref{fig-bound-probs}.
As can be seen, $\ell_{\text{SSM}}$ encounters challenges in bounding NDCG except in the upper right corner 
where the target item is ranked lower ($r_+ \ge 500$) and the number of negative samples is sufficiently large ($K \ge 300$).
This can be understood by the decreasing probabilities in sampling those hard samples ranked higher than the target item.
Consequently, the usefulness of the uniformly sampled items gradually diminishes as the model being well-trained.
In a nutshell, before considering coverage and tightness, 
necessary modifications should be made to guarantee a high bounding probability.

We would like to emphasize that a very simple modification exists without the need for sampling more items.
By scaling up the sampled normalizing term with a pre-specific weight $\alpha \ge 1$,
the resulting Scaled Cross-Entropy (SCE) becomes
\begin{equation}
    \ell_{\text{SCE}} := -s_{v_+} + \log \hat{Z}(\alpha), 
    \quad
    \hat{Z}(\alpha) = \exp(s_{v_+}) + \alpha \cdot \sum_{i=1}^K \exp(s_{v_i}).
\end{equation}
In the case of $\alpha=1$, it degrades to the normal sampled softmax loss $\ell_{\text{SSM}}$.
The following theorem describes why this scaled loss is more likely to bound NDCG.

\begin{theorem}
    \label{theorem-sce-bounds}
    Under the same conditions as stated in Theorem \ref{theorem-ssm-bounds},
    the inequality \eqref{ieq-ndcg-mrr} holds for $\ell_{\mathrm{SCE}}$ with a probability of at least
    \begin{equation}
        \label{eq-sce-bounds}
        1 - \bigl\lceil \frac{m}{\alpha} \bigr\rceil (1 - r_+ / |\mathcal{I}|)^{\lfloor K/ \lceil \frac{m}{\alpha} \rceil \rfloor }.
    \end{equation}
\end{theorem}
\begin{proof}
Proof of Theorem~\ref{theorem-sce-bounds} and corresponding results of MRR are in Appendix~\ref{section-proof-theorem12}.
\end{proof}

\begin{wrapfigure}{r}{0.45\textwidth}
	\centering
	\includegraphics[width=0.45\textwidth]{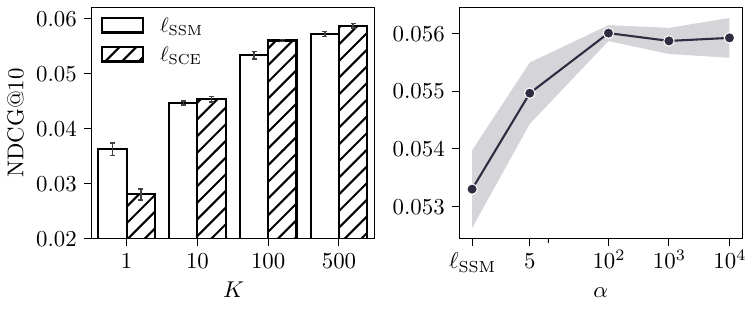}
	\caption{
        Sampled softmax loss comparison on Beauty.
        \textbf{Left}:
        $\ell_{\text{SSM}}$ versus $\ell_{\text{SCE}} \: (\alpha=100)$ across different number of negative items.
        \textbf{Right}:
        $\ell_{\text{SCE}} \: (K=100)$ with an increasing value of $\alpha$.
	}
	\label{fig-sce-ssm}
\end{wrapfigure}

As can be seen in Figure \ref{fig-bound-probs},
increasing $\alpha$ from $1$ to $5$ has already made the bounding probabilities meaningful in most areas.
Further increases guarantee a reasonable probability to bound NDCG or MRR even when the target item has been ranked highly.
In contrast to $\ell_{\text{SSM}}$, SCE theoretically requires a much smaller number of negative samples to achieve comparable recommendation performance.
Of course, SCE is not perfect:
the scaling operation inevitably raises concerns about the high variance problem.
As depicted in the left panel of Figure~\ref{fig-sce-ssm}, 
if negative samples are very rare ($K = 1$), 
a larger weight of $\alpha$ tends to worsen the ranking capability.
Fortunately, this high variance problem appears less significant 
as $K$ slightly increases (\eg, $K \ge 10$ for Beauty).
Sampling 100 negative samples for $\ell_{\text{SCE}} \: (\alpha = 100)$ produces comparable performance to using 500 negative samples for $\ell_{\text{SSM}}$ on the Beauty dataset.
Moreover, it can be inferred from the right panel that $\ell_{\text{SCE}}$ is quite robust to the choice of $\alpha$;
competitive results could be consistently obtained even with a value of $\alpha=10^4$.

\begin{figure}
	\centering
    \includegraphics[width=0.9\textwidth]{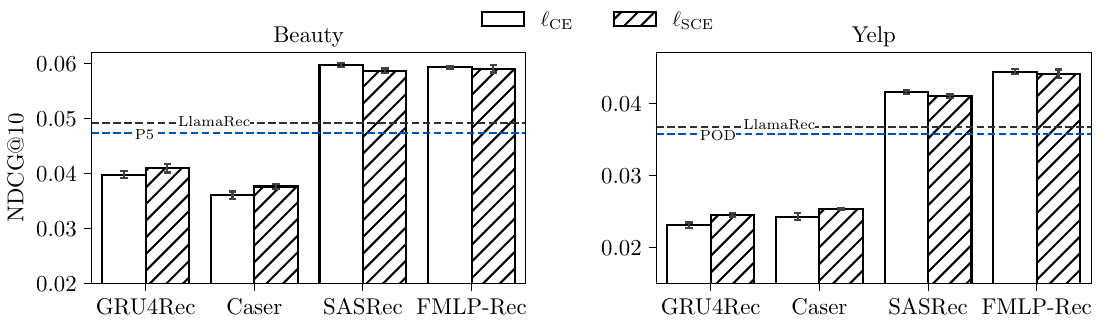}
	\caption{
        Performance comparisons between state-of-the-art LLM-based recommenders 
        and traditional models using $\ell_{\text{CE}}$ and $\ell_{\text{SCE}} \:(\alpha=100)$.
        Existing LLM-based recommenders are not as effective as claimed even in a practical scenario.
	}
    \label{fig-practical-comparisons}
\end{figure}

\textbf{LLM-based recommenders are not as effective as claimed in a practical scenario.}
For efficiency purposes, less than 5\% of all items will be sampled, specifically $K=500$ for Beauty and $K=100$ for Yelp.
Due to the sampling nature of SCE, 
compared to the 200 epochs required for CE, it sometimes consumes more iterations (300 epochs) until convergence.
Other hyper-parameters of SCE completely follow CE.
We then compare traditional models trained with SCE
to the state-of-the-art LLM-based recommenders in Figure~\ref{fig-practical-comparisons}.
As can be seen, SCE enables traditional models to achieve comparable recommendation performance by sampling only a small number of negative items.
Notably, for simple models like GRU4Rec and Caser, SCE even surpasses CE in improving ranking capability.
We hypothesize that this is because the original cross-entropy loss may be too challenging and unfocused 
for these models to optimize given their limited expressive power.
For models like SASRec and FMLP-Rec, 
the performance gains achieved by SCE are not very significant, 
but they still allow these models to outperform all other LLM-based recommenders by a substantial margin.
Considering the computational and memory costs associated with LLM, 
we believe that existing LLM-based recommenders may not be as effective as claimed in practical applications.

\section{Related Work}
\label{section-related-work}

\textbf{Recommender systems} 
are developed to enable users to quickly and accurately ascertain relevant items.
The primary principle is to learn underlying interests from user information, especially historical interactions.
Collaborative filtering \cite{loss:BPR:Rendle:2012,CF:NCF:He:2017} performs personalized recommendation 
by mapping users and items into the same latent space in which interacted pairs are close.
Beyond static user representations,
sequential recommendation \cite{shani2005mdp,li2020time} focuses on capturing dynamic interests from item sequences.
Early efforts such as GRU4Rec \cite{Seq:GRU4Rec:Hidasi:2015} and Caser~\cite{Seq:Caser:Tang:2018} 
respectively apply recurrent neural networks (RNNs) and convolutional neural networks (CNNs) to sequence modeling.
Recently, Transformer \cite{Transformer:Attention:Vaswani:2017,devlin2018bert} has become increasingly popular in recommendation 
due to its parallel efficiency and superior performance.
For example, SASRec~\cite{Seq:SASRec:Kang:2018} and BERT4Rec \cite{Seq:Bert4Rec:Sun:2019} 
respectively employ unidirectional and bidirectional self-attention.
Differently,
Zhou et al. \cite{Seq:FMLPRec:Zhou:2022} present FMLP-Rec to denoise the item sequences through learnable filters 
so that state-of-the-art performance can be obtained by mere MLP modules.

\textbf{LLM for recommendation} 
has gained a lot of attention recently because:
1) The next-token generation feature can be easily extended to the next-item recommendation (\ie, sequential recommendation);
2) The immense success of LLM in natural language processing promises the development of universal recommenders.
Some studies \cite{LLM4Rec:UncoverChatGPT:Dai:2023,LLM4Rec:Chat-Rec:Gao:2023}
have demonstrated the powerful zero/few-shot ability of LLMs (\eg, GPT \cite{LLM:GPT:Openai:2023}),
especially their potential in explainable and cross-domain recommendations \cite{LLM4Rec:CRs:Feng:2023,LLM4Rec:Chat-Rec:Gao:2023}.
Nevertheless, there is a consensus \cite{LLM4Rec:Prefrence:Kang:2023,LLM4Rec:InstructRec:Zhang:2023,LLM4Rec:TallRec:Bao:2023} 
that without domain-specific knowledge learned by fine-tuning, 
LLM-based recommenders still stay far behind traditional models.

As an early effort, 
P5~\cite{LLM4Rec:P5:Geng:2022} unifies multiple recommendation tasks into a sequence-to-sequence paradigm.
Based on the foundation model of T5~\cite{LLM:T5:Raffel:2020}, 
each task can be activated through some specific prompts.
Hua et al.~\cite{LLM4Rec:CIDIID:Hua2023} takes a further step beyond P5 
by examining the impact of various ID indexing methods,
and a combination of collaborative and independent indexing stands out.
Recently, more LLM recommenders \cite{LLM4Rec:ControlRec:Qiu:2023,LLM4Rec:LLaRA:Liao:2023,LLM4Rec:E4SRec:Li:2023,LLM4Rec:LLaMARec:Yue2023} 
based on Llama~\cite{LLM:LLama:Touvron:2023} or Llama2~\cite{LLM:LLaMA2:Touvron:2023} are developed.
For example, 
LlamaRec~\cite{LLM4Rec:LLaMARec:Yue2023} proposes a two-stage framework based on Llama2 to rerank the candidates retrieved by traditional models.
To enable LLM to correctly identify items,
E4SRec~\cite{LLM4Rec:E4SRec:Li:2023} incorporates ID embeddings trained by traditional sequence models through a linear adaptor,
and applies LORA~\cite{LLM:LORA:Hu:2021} for parameter-efficient fine-tuning.

\textbf{Cross-entropy and its approximations} \cite{loss:BPR:Rendle:2012,loss:NCE:Gutmann:2010,loss:IS:Bengio:2008,loss:SSM:Blanc:2018,loss:Consistency:Ma2018}
have been extensively studied.
The most related works are:
1) Bruch et al.~\cite{loss:Bound:Bruch:2019} theoretically connected cross-entropy to some ranking metrics;
2) Wu et al.~\cite{loss:SSM:Wu:2023} further found its desirable property in alleviating popularity bias;
and recently, 
3) Klenitskiy et al.~\cite{loss:NEG4Rec:Klenitskiy:2023} and Petrov et al.~\cite{loss:gBCE:Petrov:2023} 
respectively
applied cross-entropy and a generalized BCE loss to eliminate the performance gap between SASRec and BERT4Rec.
Differently, we are to
1) understand the superiority of cross-entropy from the tightness and coverage perspectives;
2) make some necessary modifications to the sampled softmax loss based on these findings;
and 3) facilitate an objective evaluation of LLM-based recommendation 
by acknowledging the true capability of traditional models in practical applications.

\section{Discussion and Conclusion}

The emergence of LLMs has inspired novel attempts in various fields, including recommender systems.
The great potential of explainable and cross-domain recommendations signals a promising direction toward the development of a universal recommender.
Besides, 
some researchers would like to argue that LLM-based recommenders may outperform traditional models in accuracy
by virtue of their superior world knowledge and reasoning ability.
We are skeptical of this claim because in reality, even experts find it challenging to provide personalized recommendations for different users.
The recommended items may align logically but may not accurately match a user's specific interests. 
We have empirically investigated this in Section~\ref{section-observation} and Section~\ref{section-sce},
which validates our point about existing LLM-based recommenders.
However, we also acknowledge that offline comparisons may not be entirely convincing. 
Recommendation is inherently a subjective task, and recommenders that follow logic may produce diverse recommendations to circumvent the echo chamber effect \cite{echo:Ge:2020}.

This work is not intended to discourage further exploration of the potential of LLM for personalized recommendations. 
We simply advocate a fair experimental environment to ensure that both traditional models and LLM-based recommenders can be objectively evaluated in the future. 
As demonstrated in this paper, the sampled softmax loss especially the proposed SCE can serve as a reliable criterion 
for unleashing the true ranking capability of traditional models.

\newpage

\appendix

\tableofcontents

\section{Appendices}

\subsection{Overview of Loss Function}
\label{section-loss-function}

\textbf{Cross-Entropy (CE)},
also known as the negative log-likelihood (NLL) loss, can be formulated as follows:
\begin{align*}
    \ell_{\text{CE}} 
    &= -\log \frac{\exp(s_{v_+})}{ \sum_{v \in \mathcal{I}} \exp(s_v)}
    = -s_{v_+} + \log \underbrace{\sum_{v \in \mathcal{I}} \exp(s_v)}_{=: Z_{\text{CE}}}.
\end{align*}
This is also the de facto objective commonly used in the pre-training (fine-tuning) of LLMs.

\textbf{Binary Cross-Entropy (BCE).} 
BCE samples one negative item $v_-$ for each target $v_+$,
which necessitates the recommender to possess excellent pointwise scoring capability:
\begin{align*}
    \ell_{\text{BCE}}
    &= -\log \sigma(s_{v_+}) - \log (1 - \sigma(s_{v_-}))
    = -\log \frac{\exp(s_{v_+})}{1 + \exp(s_{v_+})} 
    -\log \frac{1}{1 + \exp(s_{v_-})} \\
    &= -s_{v_+} + \log
    \underbrace{
    \big(
        (1 + \exp(s_{v_+}))
        (1 + \exp(s_{v_-}))
    \big)
    }_{=: Z_{\text{BCE}}}.
\end{align*}
Here $\sigma: \mathbb{R} \rightarrow [0, 1]$ denotes the sigmoid function.

\textbf{Bayesian Personalized Ranking (BPR) \cite{loss:BPR:Rendle:2012}}
also samples one negative item $v_-$ for each target $v_+$,
but it intends to maximize the probability that $v_+$ will be chosen in preference to $v_-$:
\begin{align*}
    \ell_{\text{BPR}} 
    &= -\log \sigma(s_{v+} - s_{v_-})
    = -\log \frac{\exp(s_{v_+})}{\exp(s_{v_+}) + \exp(s_{v_-})} \\
    &= s_{v_+} + \log
    \underbrace{
    \bigl(
        \exp(s_{v_+}) + \exp(s_{v_-})
    \bigr)
    }_{=: Z_{\text{BPR}}}.
\end{align*}

\textbf{Noise Contrastive Estimation (NCE) \cite{loss:NCE:Gutmann:2010,loss:NCE4LM:Mnih:2012}}
requires the model to discriminate the target $v_+$ from an easy-to-sample noise distribution:
\begin{align*}
    \ell_{\text{NCE}}
    &= -\log \sigma(s_{v_+}') - \sum_{i=1}^K \log (1 - \sigma(s_{v_i}')) \\
    &= -\log \frac{\exp(s_{v_+}')}{1 + \exp(s_{v_+}')} - \sum_{i=1}^K \log \frac{1}{1 + \exp(s_{v_i}')} \\
    &= -s_{v_+}' + \log 
    \underbrace{
    \biggl((1 + \exp(s_{v_+}')) \prod_{i=1}^K (1 + \exp(s_{v_i}') \biggr)
    }_{=: Z_{\text{NCE}}}.
\end{align*}
In the case of uniform sampling, $s_{v}' = s_v - c - \log \frac{K}{|\mathcal{I}|}$,
where $c$ is a trainable parameter as an estimate of $\log Z_{\text{CE}}$.

\subsection{Proof of Proposition \ref{proposition-bounding}}
\label{section-proof-prop1}

\begin{proposition}
    For a target item $v_+$ which is ranked as $r_+$, the following inequality holds true for any $n \ge r_+$
    \begin{equation}
        -\log \mathrm{NDCG}(r_+) \le -\log \mathrm{MRR}(r_+) \le  \ell_{\mathrm{CE}\text{-}n}.
    \end{equation}
    where
    \begin{equation}
        \ell_{\mathrm{CE}\text{-}n} := -s_{v_+} + \log \sum_{r(v) \le n} \exp(s_{v}).
    \end{equation}
\end{proposition}

\begin{proof}

Notice that $\log_2 (1 + x) \le x$ holds true for any $x \ge 1$. 
Hence, we have
\begin{align*}
    &\text{NDCG}(r_+) = \frac{1}{\log_2 (1 + r_+)} \ge \text{MRR}(r_+) = \frac{1}{r_+} \\
    =& \frac{1}{1 + \sum_{v \not = v_+} \delta(s_{v} > s_{v_+})}
    = \frac{1}{1 + \sum_{r(v) < r_+} \delta(s_{v} > s_{v_+})} \\
    \ge & \frac{1}{1 + \sum_{r_v < r_+} \exp(s_v - s_{v_+})} 
    = \frac{\exp(s_{v_+})}{\exp(s_{v_+}) + \sum_{r(v) < r_+} \exp(s_v)} \\
    \ge & \frac{\exp(s_{v_+})}{\sum_{r(v) \le n} \exp(s_v)}.
\end{align*}
where $\delta(\text{condition}) = 1$ if the given condition is true else 0,
and the second-to-last inequality holds because $\exp(s_v - s_{v_+}) \ge 1$ when $s_v > s_{v_+}$.
\end{proof}

\subsection{Proof of Theorem~\ref{theorem-ssm-bounds} and Theorem~\ref{theorem-sce-bounds}}
\label{section-proof-theorem12}

First, let us introduce some lemmas that give a lower bound of $\ell_{\text{SCE}}$.

\begin{lemma}
    \label{lemma-bound-scaled}
    Let $\xi$ be the number of sampled items with scores not lower than that of the target; that is
    \begin{equation}
        \xi := 
        \bigl|
            \{
                v_i: s_{v_i} \ge s_{v_+}, \: i=1,2,\ldots, K
            \}
        \bigr|.
    \end{equation}
    Then, we have
    \begin{align}
        \ell_{\mathrm{SCE}} \ge \log (1 + \alpha \xi).
    \end{align}
\end{lemma}

\begin{proof}
    According to the definition of SCE, it follows that
    \begin{align*}
        \ell_{\text{SCE}}
        &= -s_{v_+} + \log 
        \bigl(
            \exp(s_{v_+}) + \alpha \sum_{i=1}^K \exp(s_{v_i})
        \bigr) \\
        &= \log 
        \bigl(
            1 + \alpha \sum_{i=1}^K \exp(s_{v_i} - s_{v_+})
        \bigr) \\
        &\ge \log 
        \bigl(
            1 + \alpha \sum_{i=1}^K \delta(s_{v_i} \ge s_{v_+}).
        \bigr) \\
        &= \log (1 + \alpha \xi).
    \end{align*}
\end{proof}

\begin{lemma}
    \label{lemma-binomial-bound}
    Let $\xi \sim \mathcal{B}(K, p)$ denote a random variable representing the number of successes over $K$ binomial trials with a probability of $p$.
    Then, we have
    \begin{align}
        \mathbb{P}(\xi \ge m) \ge 1 - m (1 - p)^{\lfloor K / m \rfloor}, \quad \forall m=0, 1, \ldots, K.
    \end{align}
\end{lemma}

\begin{proof}\footnote{The proof follows from the response posted at the URL: \url{https://math.stackexchange.com/questions/3626472/upper-bound-on-binomial-distribution}.}
    Divide the $K$ independent binomial trials into $m$ disjoint groups, 
    each containing at least $\lfloor K/m \rfloor$ trials.
    If $\xi < m$, then one of the groups must have no successes observed; formally, we have
    \begin{align}
        \mathbb{P}(\xi < m)
        &\le \mathbb{P}(
            \bigcup_{i=1}^m \{ \text{no successes observed in group } i \} 
        ) \\
        &\le \sum_{i=1}^m \mathbb{P}(
            \{ \text{no successes observed in group } i \} 
        ) \\ 
        &\le m (1 - p)^{\lfloor K/m \rfloor}.
    \end{align}
    Hence, the proof is completed by noting the fact that
    \begin{align}
        \mathbb{P}(\xi \ge m) = 1 - \mathbb{P}(\xi < m).
    \end{align}
\end{proof}

Now we are ready to prove Theorem~\ref{theorem-ssm-bounds} and Theorem~\ref{theorem-sce-bounds},
which directly follow from the subsequent theorem.

\begin{theorem}
    Let $v_+$ be a target item which is ranked as $r_+ \le 2^m - 1$ for some $m \in \mathbb{N}$,
    If we uniformly sample $K$ items for training, then we have
    \begin{align}
        \mathbb{P}
        \bigl(
            -\log \mathrm{NDCG}(r_+) \le \ell_{\mathrm{SCE}}
        \bigr)
        \ge 1 - \bigl\lceil \frac{m}{\alpha} \bigr\rceil (1 - r_+ / |\mathcal{I}|)^{\lfloor K/ \lceil \frac{m}{\alpha} \rceil \rfloor }, \\
        \mathbb{P}
        \bigl(
            -\log \mathrm{MRR}(r_+) \le \ell_{\mathrm{SCE}}
        \bigr)
        \ge 1 - \bigl\lceil \frac{2^m}{\alpha} \bigr\rceil (1 - r_+ / |\mathcal{I}|)^{\lfloor K/ \lceil \frac{2^m}{\alpha} \rceil \rfloor }.
    \end{align}
\end{theorem}

\begin{proof}

As $r_+ \le 2^m - 1$ for some $m \in \mathbb{N}$, we immediately have
\begin{equation*}
    -\log \text{NDCG}(r_+) \le   \log m, 
    \quad
    -\log \text{MRR}(r_+) \le   m \log 2. 
\end{equation*}

    Lemma \ref{lemma-bound-scaled} implies that
    \begin{align}
        \mathbb{P}
        \bigl(
            -\log \mathrm{NDCG} \le \ell_{\text{SCE}}
        \bigr)
        & \ge 
        \mathbb{P}
        \bigl(
            \log (1 + \alpha \xi) \ge \log m
        \bigr) \\
        &=
        \mathbb{P}
        \bigl(
            \xi \ge \frac{m - 1}{\alpha}
        \bigr)
        \ge
        \mathbb{P}
        \bigl(
            \xi \ge \frac{m}{\alpha}
        \bigr) \\
        &=
        \mathbb{P}
        \bigl(
            \xi \ge \bigl\lceil \frac{m}{\alpha} \bigr\rceil
        \bigr).
    \end{align}
    The last equality holds because $\xi$ is an integer random variable.
    Also notice that uniformly sampling from $\mathcal{I}$ yields a hit probability of $p=r_+ / |\mathcal{I}|$ such that
    the score of the sampled item is not lower than that of the target (\ie, the top-$r_+$ ranked items).
    Therefore, based on Lemma \ref{lemma-binomial-bound}, we have
    \begin{align}
        \mathbb{P}
        \bigl(
            \xi \ge \bigl\lceil \frac{m}{\alpha} \bigr\rceil
        \bigr)
        &\ge 1 - \bigl\lceil \frac{m}{\alpha} \bigr\rceil (1 - r_+ / |\mathcal{I}|)^{\lfloor K/ \lceil \frac{m}{\alpha} \rceil \rfloor }.
    \end{align}
    The proof for MRR is analogous by noting that
    \begin{align}
        \mathbb{P}
        \bigl(
            -\log \mathrm{MRR} \le \ell_{\text{SCE}}
        \bigr)
        & \ge 
        \mathbb{P}
        \bigl(
            \log (1 + \alpha \xi) \ge m \log 2
        \bigr) \\
        &=
        \mathbb{P}
        \bigl(
            \xi \ge \frac{2^m - 1}{\alpha}
        \bigr)
        \ge
        \mathbb{P}
        \bigl(
            \xi \ge \frac{2^m}{\alpha}
        \bigr) \\
        &=
        \mathbb{P}
        \bigl(
            \xi \ge \bigl\lceil \frac{2^m}{\alpha} \bigr\rceil
        \bigr).
    \end{align}

\end{proof}

\subsection{Datasets}
\label{section-dataset}

\begin{table}[h]
  \centering
  \caption{Dataset statistics.}
  \label{table-dataset-statistics}
  \scalebox{0.8}{
\begin{tabular}{c|ccccc}
  \toprule
Dataset     & \#Users & \#Items & \#Interactions & Density & Avg. Len. \\
  \midrule
Beauty      & 22,363  & 12,101  & 198,502        & 0.07\%  & 8.9       \\
Yelp       & 30,431  & 20,033  & 316,354        & 0.05\%  & 10.4       \\
  \bottomrule
\end{tabular}
  }
\end{table}

In this study, we perform experiments on two public datasets. 
Specifically, the Beauty dataset is extracted from Amazon reviews\footnote{
\url{https://cseweb.ucsd.edu/~jmcauley/datasets/amazon/links.html}
}, 
while the Yelp\footnote{
\url{https://www.yelp.com/dataset}
} dataset collects the interactions that occured in 2019.

\subsection{Baselines}
\label{section-baseline}

\begin{itemize}[leftmargin=*]
    \item \textbf{P5 (CID+IID)}~\cite{LLM4Rec:CIDIID:Hua2023} 
    unifies multiple tasks (\eg, sequential recommendation and rating prediction) into a sequence-to-sequence paradigm.
    The use of collaborative and independent indexing together creates LLM-compatible item IDs.
    \item \textbf{POD}~\cite{LLM4Rec:POD:Li2023} 
    bridges IDs and words by distilling long discrete prompts into a few continuous prompts.
    It also suggests a task-alternated training strategy for efficiency.
    \item \textbf{LlamaRec}~\cite{LLM4Rec:LLaMARec:Yue2023}
    aims to address the slow inference process caused by autoregressive generation.
    Given the candidates retrieved by traditional models, it subsequently reranks them based on the foundation model of Llama2.
    \item \textbf{E4SRec}~\cite{LLM4Rec:E4SRec:Li:2023}
    incorporates ID embeddings
    trained by traditional sequence models through a linear adaptor,
    and applies LORA \cite{LLM:LORA:Hu:2021} for parameter-efficient fine-tuning.
\end{itemize}
Additionally, 
five sequence models, covering MLP, CNN, RNN, and Transformer architectures,
are considered here to
uncover the true capability of traditional methods.
\begin{itemize}[leftmargin=*]
    \item \textbf{GRU4Rec} \cite{Seq:GRU4Rec:Hidasi:2015} 
    applies RNN to recommendation with specific modifications made to cope with data sparsity.
    \item \textbf{Caser} \cite{Seq:Caser:Tang:2018} treats the embedding matrix as an `image',
    and captures local patterns by utilizing traditional filters.
    \item \textbf{SASRec} \cite{Seq:SASRec:Kang:2018}
    is a pioneering work equipped with a unidirectional self-attention mechanism.
    By its nature, SASRec predicts the next item based on previously interacted items.
    \item \textbf{FMLP-Rec} \cite{Seq:FMLPRec:Zhou:2022} denoises item sequences in the frequency domain.
    Although FMLP-Rec consists solely of MLPs, it performs better than Transformer-based models.
\end{itemize}

\subsection{Computational Complexity}
\label{section-time}

The major cost of cross-entropy lies in the inner product and softmax operations.
It has a complexity of $\mathcal{O}(|\mathcal{I}|d)$, where $d$ denotes the embedding size before similarity calculation.
In contrast, the approximations require a lower cost $\mathcal{O}(Kd)$, 
correspondingly an additional overhead $\mathcal{O}(K)$ for uniform sampling.
Overall, it is profitable if $K \ll |\mathcal{I}|$.
The experiments of traditional models are conducted on an Intel Xeon CPU E5-2680 v4 platform and a single RTX 3090 GPU,
while those for LLM-based recommenders are performed on an Intel Xeon Gold 6133 CPU platform and four A100 GPUs.

\subsection{Detailed Settings}
\label{section-settings}

For traditional models considered in this paper, 
we use the Xavier/Normal initialization method to randomly initialize the embedding table of dimension 64.
The Adam optimizer is employed for training, whose learning rate and weight decay are retuned in the range of \{1e-4, 5e-4, 1e-3, 5e-3\} and [0, 5e-4], respectively.
200 epochs is enough for CE while sometimes 300 epochs are required for BCE, BPR, and SCE.
For LLM-based recommenders, we conduct experiments following their source code and paper settings.
In particular, the learning rate for POD and P5 are returned within \{1e-5, 5e-5, 1e-4, 5e-4, 1e-3\}.

\subsection{Broader Impact}
\label{section-broader-impact}

LLM-based recommenders still suffer from enormous computational burden,
the observations that existing LLM-based recommenders are not as effective as claimed may prevent them from practical applications,
so as to avoid wastage of resources.
On the other hand, this work may slow down the progress of LLM-based recommenders in personalized recommendations,
although this is not our intention.
We simply advocate a fair experimental environment to ensure that both traditional models and LLM-based recommenders can be objectively evaluated in the future.

\subsection{Limitations}
\label{section-limitations}

There are three major limitations.
Firstly, the offline comparisons may not be entirely convincing. 
Recommendation is inherently a subjective task, and recommenders that follow logic may produce diverse recommendations to circumvent the echo chamber effect \cite{echo:Ge:2020}.
Secondly, 
as we notice that the next-token generation feature of LLMs can be easily extend to the next-item recommendation,
only sequential recommendation is considered in this paper.
Hence, the observations and conclusions presented in this study may differ when applied to general recommendation \cite{CF:NCF:He:2017,CF:LightGCN:He:2020}.
Finally, the proposed SCE encounters a high variance problem when the negative items are very rare ($K=1$).
Fortunately, for the Beauty dataset, a value of $K \ge 10$ is sufficient to achieve satisfactory results, 
which is computationally friendly in practice.

\end{document}